Article

# The Transcription Factor E4F1 Coordinates CHK1-Dependent Checkpoint and Mitochondrial Functions

## Graphical Abstract

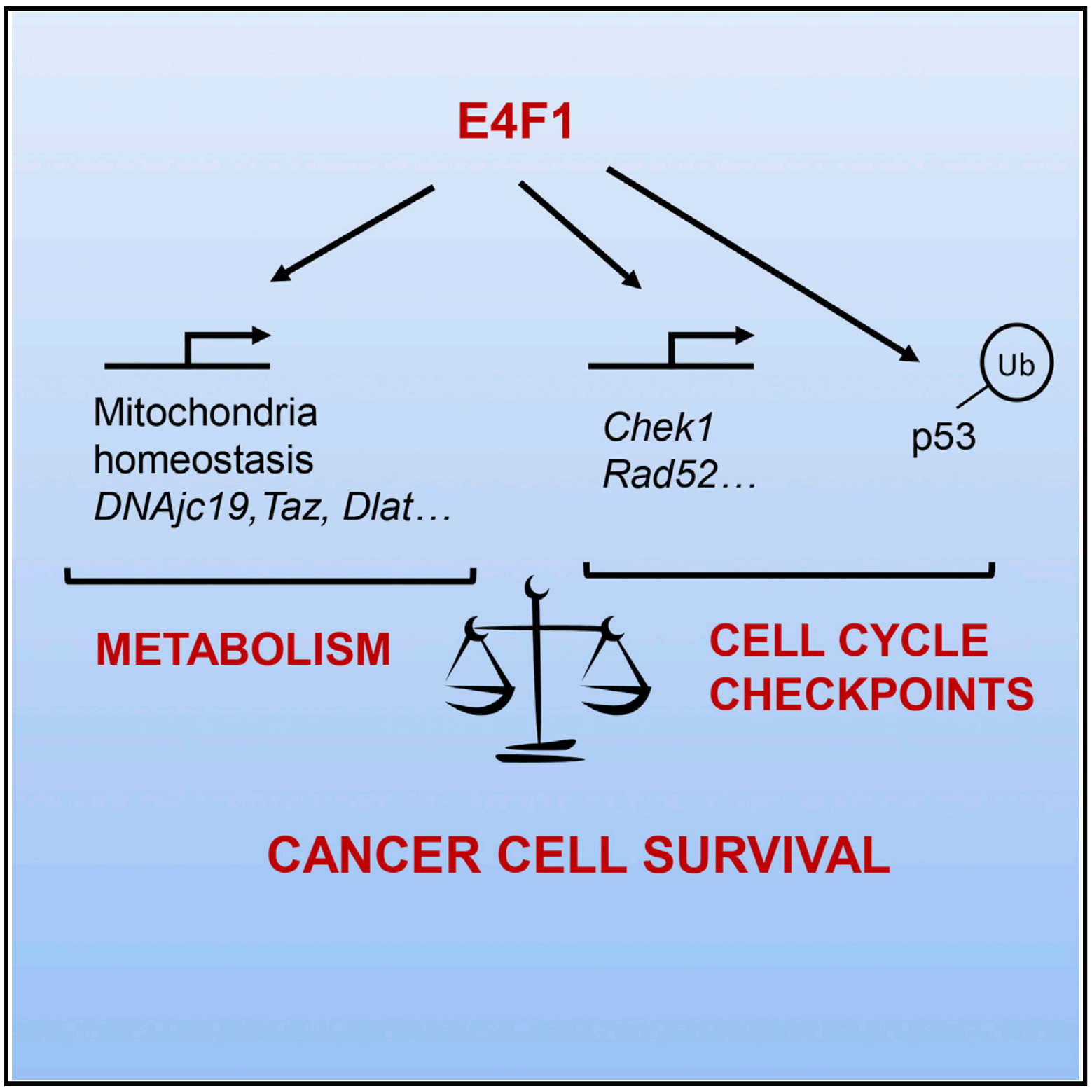

## Authors

Geneviève Rodier, Olivier Kirsh, ..., Laurent Le Cam, Claude Sardet

## Correspondence

genevieve.rodier@inserm.fr (G.R.),
laurent.lecam@inserm.fr (L.L.C.),
claude.sardet@inserm.fr (C.S.)

## In Brief

Rodier et al. show that the transcription factor E4F1, previously described as a regulator of the p53 response and a target of E1A, directly controls a transcriptional program involved in mitochondrial homeostasis, metabolism, and cell-cycle checkpoints. This program is essential for cancer cell survival.

## Highlights

- Transcriptional target genes of E4F1 encode mitochondrial and checkpoint proteins
- E4F1 controls the basal expression of the *Chek1* gene
- E4F1 transcriptional program impacts metabolism and stress response
- E4F1 is essential for cancer cell survival and could be a target for cancer therapies

## Accession Numbers

GSE57242

# The Transcription Factor E4F1 Coordinates CHK1-Dependent Checkpoint and Mitochondrial Functions

Geneviève Rodier,[1,2,3,4,5,9,*] Olivier Kirsh,[1,4,9,11] Martín Baraibar,[6] Thibault Houlès,[1,2,3,4,5] Matthieu Lacroix,[2,3,4,5] Hélène Delpech,[1,2,3,4,5] Elodie Hatchi,[1,4] Stéphanie Arnould,[1,2,3,4,5] Dany Severac,[7] Emeric Dubois,[7] Julie Caramel,[1,4,12] Eric Julien,[1,2,3,4,5] Bertrand Friguet,[6,8] Laurent Le Cam,[2,3,4,5,10,*] and Claude Sardet[1,2,3,4,5,10,*]

[1]Equipe labellisée Ligue Contre le Cancer, Institut de Génétique Moléculaire de Montpellier, UMR5535, Centre National de la Recherche Scientifique (CNRS), 34293 Montpellier, France
[2]Institut de Recherche en Cancérologie de Montpellier (IRCM), 34298 Montpellier, France
[3]Institut National de la Santé et de la Recherche Médicale (INSERM), U1194, 34298 Montpellier, France
[4]Université de Montpellier, 34090 Montpellier, France
[5]Institut régional du Cancer de Montpellier, 34298 Montpellier, France
[6]Sorbonne Universités, UPMC Univ Paris 06, UMR 8256, Biological Adaptation and Ageing—IBPS, 75005 Paris, France
[7]MGX-Montpellier GenomiX, c/o Institut de Génomique Fonctionnelle, 141 rue de la cardonille, 34094 Cedex 5 Montpellier, France
[8]INSERM U1164, 75005 Paris, France
[9]Co-first author
[10]Co-senior author
[11]Present address: Epigenetics and Cell Fate, University Paris Diderot, Sorbonne Paris Cité, UMR7216 CNRS, 75013 Paris, France
[12]Present address: Inserm UMR-S1052, CNRS UMR5286, Centre de Recherche en Cancérologie de Lyon, 69008 Lyon, France
*Correspondence: genevieve.rodier@inserm.fr (G.R.), laurent.lecam@inserm.fr (L.L.C.), claude.sardet@inserm.fr (C.S.)

## SUMMARY

**Recent data support the notion that a group of key transcriptional regulators involved in tumorigenesis, including MYC, p53, E2F1, and BMI1, share an intriguing capacity to simultaneously regulate metabolism and cell cycle. Here, we show that another factor, the multifunctional protein E4F1, directly controls genes involved in mitochondria functions and cell-cycle checkpoints, including *Chek1*, a major component of the DNA damage response. Coordination of these cellular functions by E4F1 appears essential for the survival of p53-deficient transformed cells. Acute inactivation of *E4F1* in these cells results in CHK1-dependent checkpoint deficiency and multiple mitochondrial dysfunctions that lead to increased ROS production, energy stress, and inhibition of de novo pyrimidine synthesis. This deadly cocktail leads to the accumulation of uncompensated oxidative damage to proteins and extensive DNA damage, ending in cell death. This supports the rationale of therapeutic strategies simultaneously targeting mitochondria and CHK1 for selective killing of p53-deficient cancer cells.**

## INTRODUCTION

E4F1 transcription factor was found to regulate the viral E4 and E1A promoters (Lee and Green, 1987; Fernandes and Rooney, 1997), but its cellular target genes remain largely unknown. Characterization of E4F1 protein indicated that, in addition to its intrinsic transcriptional activities, E4F1 is also an atypical ubiquitin E3 ligase for the p53 tumor suppressor. Surprisingly, E4F1-mediated ubiquitylation does not control p53 protein stability, but regulates its transcription functions (Le Cam et al., 2006). The central role of E4F1 in the p53 pathway also is underscored by its direct interactions with p14$^{ARF}$ (Rizos et al., 2003), the polycomb member BMI1 (Chagraoui et al., 2006), and the product of the p53 target gene FHL2 (Paul et al., 2006). Compelling evidence supports the implication of E4F1 in carcinogenesis as illustrated by its interactions with pRB (Fajas et al., 2000), RASSF1A (Fenton et al., 2004), HNF1 (Dudziak et al., 2008), HMGA2 (Tessari et al., 2003), Smad4 (Nojima et al., 2010), and TCF3 (Ro and Dawid, 2011). Through its multiple activities, the E4F1 protein plays key roles in cell proliferation and survival of somatic, stem, and cancer cells, and is essential for both embryonic development and adult tissue homeostasis (Chagraoui et al., 2006; Fajas et al., 2000, 2001; Hatchi et al., 2011; Lacroix et al., 2010; Le Cam et al., 2004, 2006; Paul et al., 2006).

Here we show that the multifunctional protein E4F1 directly controls on one hand mitochondria homeostasis and metabolic functions, and on the other hand the CHK1-dependent checkpoint, both functions being required for cell survival of cancer cells. These data provide a molecular explanation for the essential role of E4F1 during cell-cycle control and cell survival and open new questions related to its role in metabolism.

## RESULTS

### E4F1 Is Necessary for the Survival of *p53$^{KO}$* and Transformed MEFs

Our previous data showed that *E4F1* inactivation induces massive cell death in *cdkn2a$^{-/-}$* myeloid tumor cells, suggesting

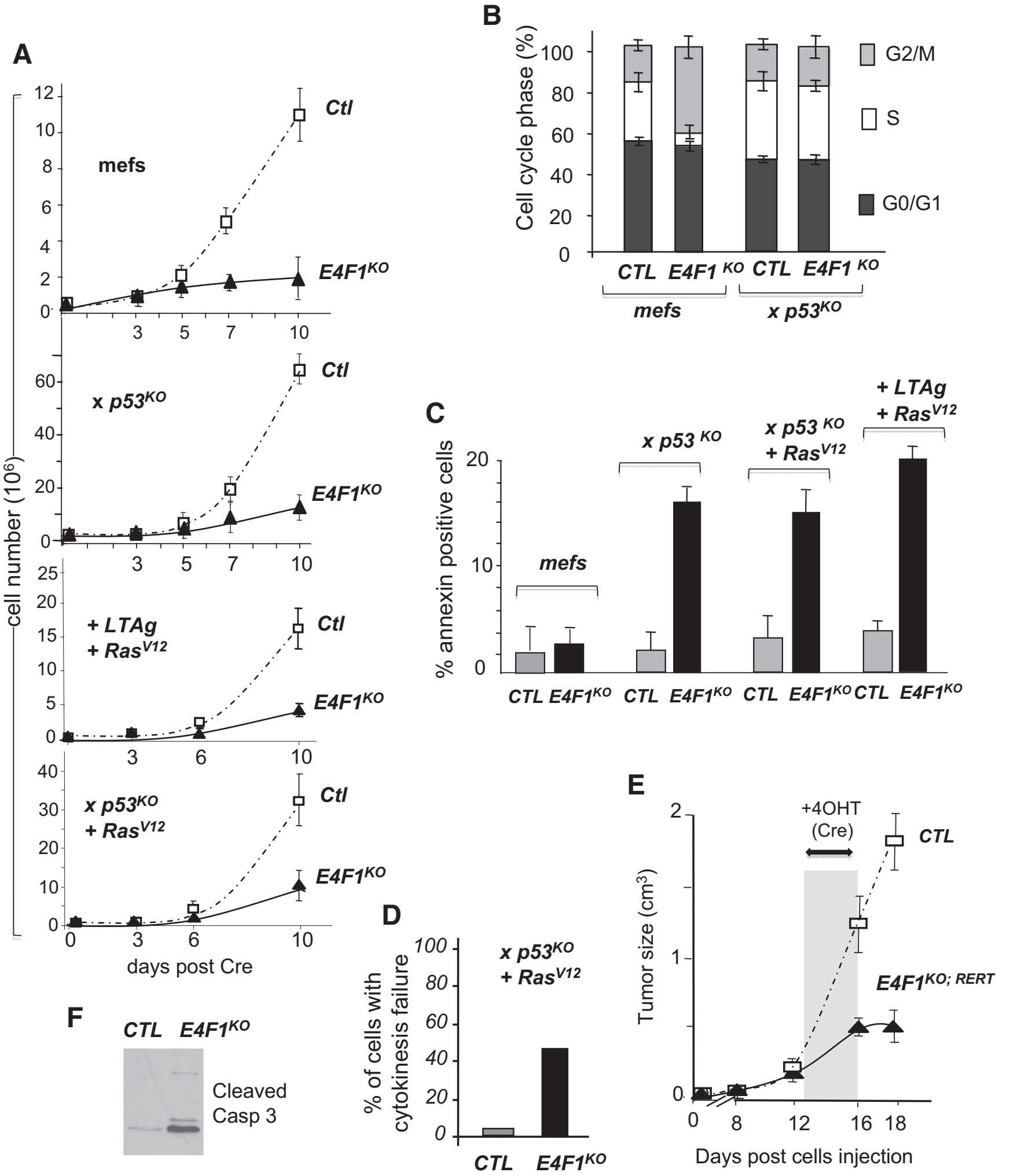

**Figure 1. E4F1 Is Essential for the Survival of $p53^{KO}$ and Transformed MEFs**

(A) $E4F1^{KO}$ cells show proliferation defects. Proliferation curves upon Cre-mediated acute inactivation of *E4F1* in $E4F1^{+/f}$ (CTL) and $E4F1^{-/f}$ ($E4F1^{KO}$) primary MEFs (first), in isogenic $p53^{KO}$ primary MEFs (second), in MEFs transformed by LTAg + $HaRas^{V12}$ (third), and in $p53^{KO}$ MEFs transformed by $HaRas^{V12}$ alone (fourth) are shown. Data represent the mean ± SEM of six experiments.

(B) *E4F1* inactivation in primary MEFs leads to p53-dependent growth arrest. Cell-cycle profiles were analyzed by FACScan (BrdU/propidium iodide [PI] stainings) 5 days after Cre-mediated inactivation of *E4F1*. Data represent the mean ± SEM of four experiments.

(C) *E4F1* inactivation in transformed cells leads to cell death. FACScan analysis of annexin-FITC-positive cells, 6 days after Cre-mediated inactivation of *E4F1*, is shown. Data represent the mean ± SEM of six experiments.

(D) Fraction of the cell population of transformed MEFs harboring abnormal mitosis upon Cre-mediated inactivation of *E4F1*, as monitored by videomicroscopy on living cells (n = 3), is shown.

(E) Acute inactivation of *E4F1* impairs tumor growth in vivo. Tumor growth (n = 3) in nude mice upon injection of *RERT*;$E4F1^{+/f}$ (CTL) and *RERT*;$E4F1^{-/f}$ MEFs, both fully transformed by LTAg + $HaRas^{V12}$, is shown. In measurable tumors, acute inactivation of *E4F1* was induced by topical skin application of tamoxifen (3 mg/mouse/day during 3 days, gray bar). Data represent the mean ± SEM of six experiments.

(F) Immunoblot analysis of cleaved-caspase 3 in protein extracts prepared from tumors, upon tamoxifen-induced in vivo inactivation of *E4F1*, is shown.

(See also Figure S1.)

that E4F1 is required for the survival of tumor cells deficient for the p53 pathways (Hatchi et al., 2011).

In this report, we compared the effect of the conditional inactivation of *E4F1* in normal, $p53^{-/-}$($p53^{KO}$), and transformed mouse embryonic fibroblasts (MEFs). Primary MEFs were derived from $E4F1^{-/flox}$ and $p53^{KO}$;$E4F1^{-/flox}$ embryos bearing a null and conditional *E4F1* alleles (Caramel et al., 2011; Hatchi et al., 2011; Lacroix et al., 2010). $E4F1^{-/flox}$ or control $E4F1^{+/flox}$ MEFs were fully transformed by *SV40-LargeT* (*LTAg*) or *E1A12S* oncogenes in combination with $Ha\text{-}Ras^{V12}$, while $p53^{KO}$;$E4F1^{-/flox}$ or control $p53^{KO}$;$E4F1^{+/flox}$ MEFs were transformed by $Ha\text{-}Ras^{V12}$ alone. Efficient depletion of E4F1 protein was obtained in these primary or transformed MEFs upon transduction with a self-excising retroviral vector encoding the Cre recombinase (Silver and Livingston, 2001; Figures S1A and S1B).

Upon *E4F1* inactivation, primary MEFs (Cre-treated $E4F1^{-/flox}$ cells, hereinafter referred to as $E4F1^{KO}$ cells) gradually ceased proliferating (Figure 1A), failed to incorporate BrdU (Figure S1D), and showed a total DNA content indicative of a cell-cycle arrest in G0–G1 and G2/M phases (Figure 1B). Importantly, $E4F1^{KO}$-arrested MEFs showed no sign of increased cell death (annexin-V-positive cells, Figure 1C) or of premature senescence (SA β-galactosidase-positive cells, Figure S1F). $E4F1^{KO}$ cells undergo a p53-dependent G2 arrest (Taylor et al., 1999) rather than an M-phase arrest, as shown by their 4n DNA content and concomitant absence of phospho-histone H3 staining and (Figure S1E) lack of CYCLIN B1 and CDK1 expression (Figure S1H). This G2/M arrest was not detected in $E4F1^{KO}$;$p53^{KO}$ double knockout (KO) cells nor in $E4F1^{KO}$ transformed cells (*LTAg*;$Ha\text{-}Ras^{V12}$ or $p53^{KO}$;$Ha\text{-}Ras^{V12}$ MEFs) (Figures 1B, S1D, and S1H). However, their growth curves (Figures 1A and S1C) and microscopic observation (Figure S1M) indicated they still exhibited severe proliferation defects compared to control cells (Figures 1A, S1C, and S1M). Time-lapse imaging confirmed that these cells died during mitosis, following one or several unsuccessful attempts to divide (Figures 1D and S1N). This resulted in a gradual increase in cell death, as indicated by annexin-V labeling (Figures 1C and S1I) and by the presence of a large number of floating dying cells with a sub-G1 DNA content (Figure 5F). These cells also released cytochrome *c* from mitochondria (Figure S1L), suggesting that they died by apoptosis.

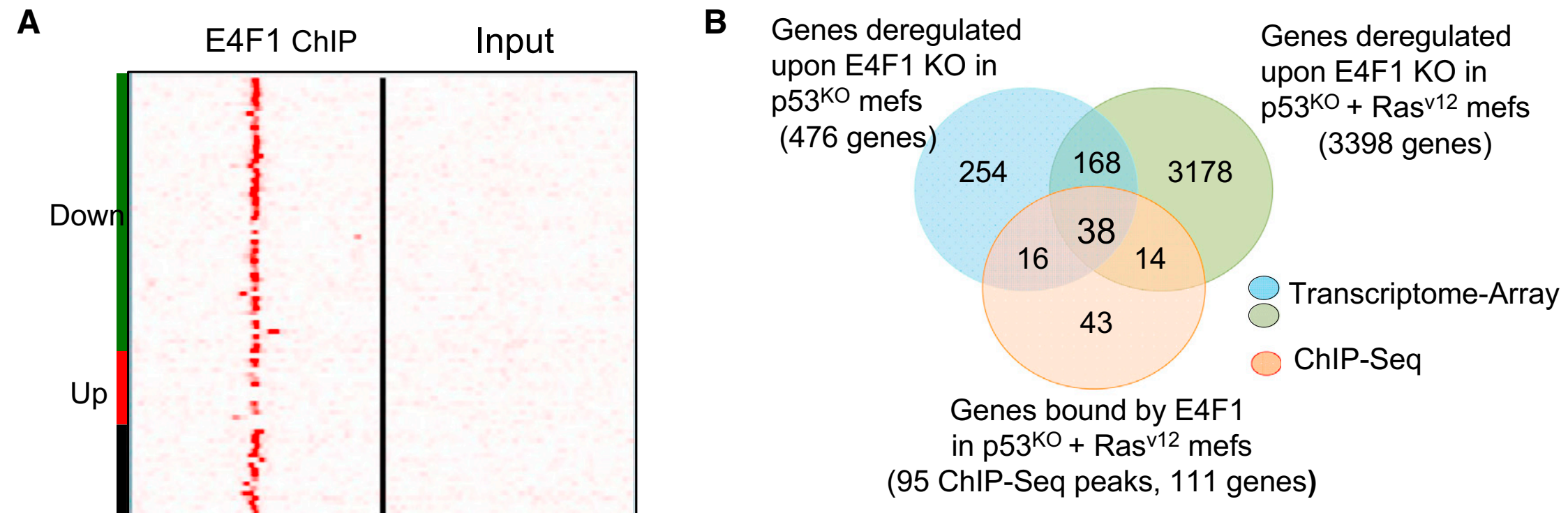

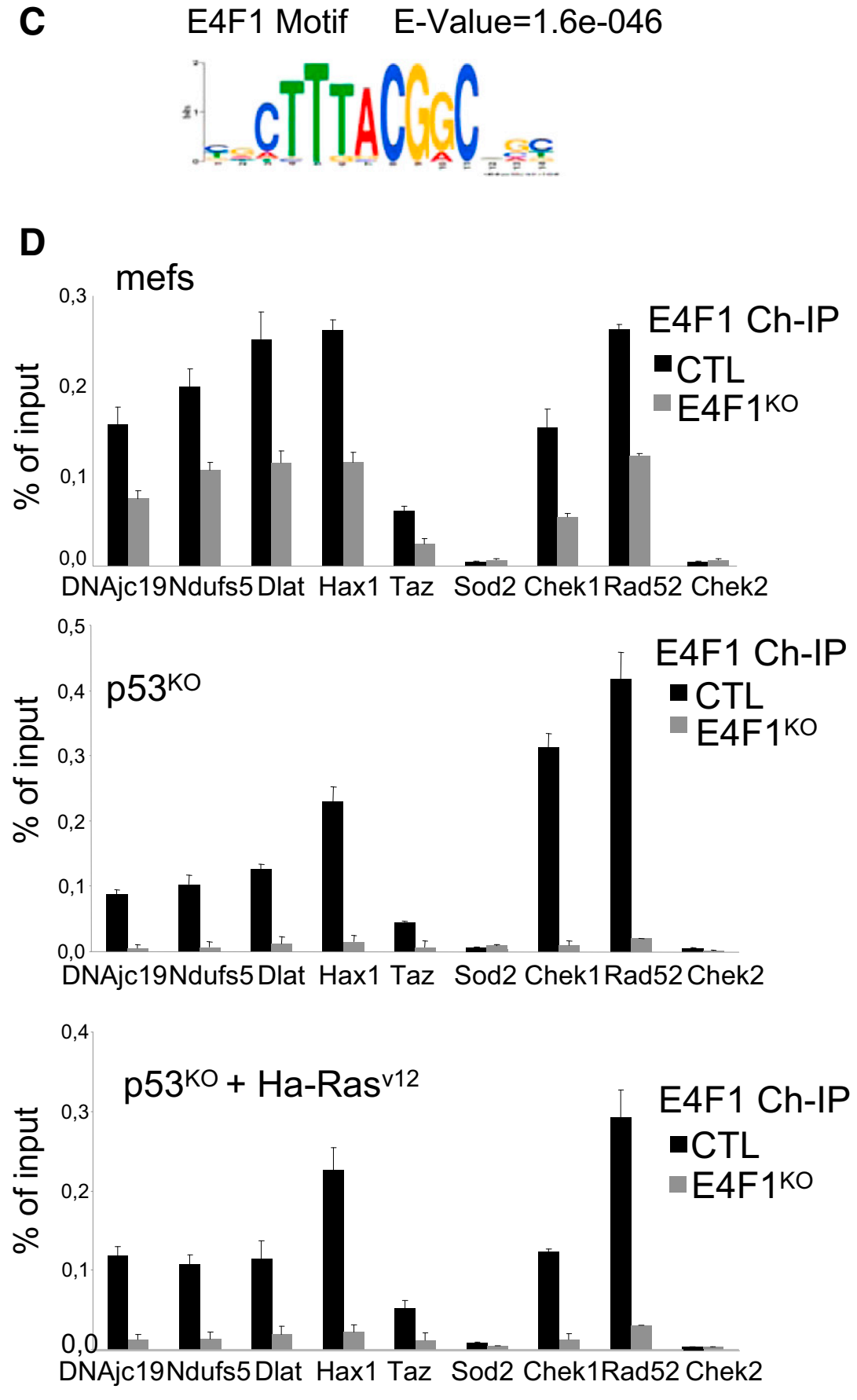

E

| Gene | Ratio KO/WT | Pos. to TSS | Motif | Function |
|---|---|---|---|---|
| *Dnajc19* | 0,06 | 74 | + | Mito |
| *Ndufs5* | 0,11 | -45 | + | Mito |
| *Mrfap1* | 0,14 | -106 | + | |
| *1700025K23Rik* | 0,18 | 352 | - | |
| *Psmb6* | 0,18 | 16 | + | |
| *Ipo8* | 0,2 | -60 | + | |
| *Senp8* | 0,23 | -168 | + | DDR/Chk |
| *Dlat* | 0,24 | 161 | + | Mito |
| *E4f1* | 0,25 | -18 | + | |
| *Taz* | 0,25 | 164 | + | Mito |
| *Tsen2* | 0,26 | 53 | + | |
| *Elp3* | 0,27 | -102 | + | |
| *Pbrm1* | 0,27 | 76 | + | |
| *Hax1* | 0,31 | -91 | + | Mito |
| *Mrpl15* | 0,32 | -67 | + | Mito |
| *Ncoa5* | 0,32 | -195 | + | |
| *Brp44l* | 0,33 | -128 | + | Mito |
| *Rad52* | 0,33 | 34 | + | DDR/Chk |
| *Dnajc9* | 0,34 | -67 | + | |
| *Eef1g* | 0,36 | 144 | + | |
| *Fktn* | 0,42 | 46 | + | |
| *AA408296* | 0,43 | -79 | + | |
| *Bag5* | 0,43 | -107 | + | |
| *Tube1* | 0,43 | 21 | + | |
| *Chek1* | 0,44 | -56 | + | DDR/Chk |
| *Tgds* | 0,47 | -112 | + | |
| *4930402E16Rik* | 0,48 | -45 | + | Mito |
| *Wdr7* | 0,49 | 51 | + | |
| *BC019943/Tti2* | 0,52 | -38 | + | DDR/Chk |
| *D1Bwg0212e* | 0,52 | 28 | + | |
| *Plod1** | 0,52 | -4189 | + | |
| *Taf5* | 0,61 | 12 | + | |
| *2510003E04Rik* | 0,63 | -94 | + | Mito |
| *2510039O18Rik** | 0,64 | 71 | + | |
| *2310003H01Rik* | 0,65 | -114 | + | DDR/Chk |
| *Mcts2* | 0,69 | -9 | + | |
| *Lsm4* | 2,54 | 95 | + | |
| *Cep290* | 2,67 | -598 | + | |

**Figure 2. E4F1 Controls a Limited Set of Genes that Define at Least Two Subprograms Involved in Mitochondrial Homeostasis and Metabolism and in Stress Response and Checkpoints**

(A) E4F1-binding sites identified by ChIP-seq are mainly located nearby TSS. E4F1 ChIP-seq read densities in the 10-kb regions surrounding the TSS of the closest genes located nearby E4F1-binding sites, as identified in *p53^KO^;HaRas^V12^* transformed MEFs, are shown.

(B) Venn diagram of the overlaps among all E4F1-bound genes identified by ChIP-seq in $p53^{KO}$;$HaRas^{V12}$ transformed MEFs (GSE57228, in red) and genes differentially expressed (transcriptiome arrays GSE57240) between $E4F1^{WT}$ and $E4F1^{KO}$ cells ($p53^{KO}$;$E4F1^{KO}$ versus $p53^{KO}$;$E4F1^{WT}$ primary MEFs [blue] and $p53^{KO}$;$HaRas^{V12}$;$E4F1^{KO}$ versus $p53^{KO}$;$HaRas^{V12}$;$E4F1^{WT}$ transformed MEFs [green]) is shown. These genome-wide analyses were performed at a time when

(legend continued on next page)

Next, we assessed the capacity of $E4F1^{WT}$ and $E4F1^{KO}$ cells to grow in soft agar and form tumors in nude mice. To bypass the usual low efficiency of infection by Cre-retroviruses in these assays, we performed these experiments on *LTAg* and *Ha-Ras*$^{V12}$ transformed MEFs derived from E4F1$^{+/flox}$;*Cre-ER*$^{T2\ KI/KI}$ and E4F1$^{-/flox}$;*Cre-ER*$^{T2\ KI/KI}$ mice that express a 4-hydroxy-tamoxifen (4OHT)-inducible CreER$^{T2}$ recombinase (Lacroix et al., 2010). 4-OHT-mediated acute inactivation of *E4F1* in these transformed cells led to a strong reduction of their capacity to form colonies in soft agar (Figure S1J) and strongly decreased their tumorigenic potential in nude mice (Figures 1E and S1K).

Altogether, these data indicate that E4F1 depletion leads to a p53-dependent cell-cycle arrest in normal MEFs and to p53-independent mitotic cell death in immortalized or transformed cells deficient for the p53 pathway.

## E4F1 Target Genes, a Limited Set of Genes Involved in Mitochondrial Homeostasis, Metabolism, and Checkpoints

E4F1 chromatin immunoprecipitation (ChIP) combined with deep sequencing (ChIP-seq) and differential transcriptomic analyses of $E4F1^{KO}$ and $E4F1^{WT}$ cells were performed to identify the p53-independent program directly controlled by E4F1. We identified the full repertoire of endogenous target DNA sites bound by E4F1 in primary MEFs and $p53^{KO}$;*Ha-Ras*$^{V12}$ transformed MEFs (Table S1) by ChIP-seq using a previously characterized anti-E4F1 rabbit polyclonal antibody (Fajas et al., 2000; Figure S1A). E4F1-binding regions were defined by combining bioinformatic toolboxes provided by CisGenome and Qeseq software systems. Raw sequencing data are available on Gene Expression Omnibus (GEO) dataset repository (GSE57228). Genome sequences significantly enriched (Table S1) by this E4F1 ChIP-seq were then annotated according to the transcription start site (TSS) of the closest gene (Table S2). Peak read densities, displayed as a heatmap (Seqminer 1.3.3) of the 10-kb (Figure 2A) or 100-kb window (Figure S2C), centered on gene TSS. Our analyses identified 126 E4F1-binding sites that are mainly located in the promoter (<1,000 bp from TSS), 5′UTR, or first exons (Figure S2B) of 111 genes that are distributed along all chromosomes (Figure S2A). Strikingly, this ChIP-seq profile was very similar in primary and $p53^{K0}$;*Ha-Ras*$^{V12}$ MEFs, indicating this E4F1 cistrome is independent of p53 and of cell transformation (Figure S2F).

In parallel, we performed differential transcriptomic analyses (microarray GEO dataset GSE57240) on $E4F1^{WT}$;$p53^{KO}$ and $E4F1^{KO}$;$p53^{KO}$ cells or their *Ha-Ras*$^{V12}$ transformed counterparts. Gene lists identified by these transcriptomic analyses were intersected with the list of genes directly bound by E4F1 according to our ChiP-seq (Figure 2B). This led to the identification of a common set of 38 genes, the expression of which depends on the presence of E4F1 nearby or on their TSS, in all tested cell populations. Notably, 36 of these 38 genes were downregulated upon *E4F1* inactivation, suggesting that E4F1 acts mainly as a transactivator for its target genes (Figure 2E).

A set of 26 genes was further validated as direct targets for E4F1 by ChIP-qPCR experiments performed in $E4F1^{WT}$ and $E4F1^{KO}$ MEFs. These assays confirmed the presence of the E4F1 protein on the TSS region of these genes in $E4F1^{WT}$, but not in $E4F1^{KO}$ MEFs (Figures 2D and 2E; Table S2). A selection of E4F1 genes was further validated by ChIP-qPCR in primary, immortalized ($p53^{KO}$) and transformed ($p53^{KO}$;*Ha-Ras*$^{V12}$) MEFs (Figure 2D).

Analyses of DNA sequences bound by E4F1 unveiled a hitherto unidentified consensus motif (Figure 2C) present in most of these target sequences (Figure 2E). Noteworthy, this new E4F1 motif was detected in many but not all of the 128 DNA regions initially identified by ChIP-seq, with an enrichment in regions located in the close vicinity of TSS (Table S2). In contrast, most peaks located further than 1 kb relative to the TSS did not harbor such an E4F1 motif and turned out to be false positive in our ChIP-qPCR validation experiments performed on $E4F1^{WT}$ and $E4F1^{KO}$ cells (Table S2). Finally, we confirmed by electrophoretic mobility shift assays (EMSAs) that this new motif is recognized by recombinant E4F1 protein in vitro (Figure S2D). Interestingly, the E4F1 gene itself contains this consensus motif nearby its own TSS and was identified as a direct target of the E4F1 protein by our ChIP-seq (Figure 2E), indicating that E4F1 controls its own expression.

Ontology analysis of this list of 38 genes, bound and transcriptionally controlled by E4F1, revealed an unexpected enrichment for nuclear genes encoding mitochondrial proteins (*Dnajc19*, *Taz*, *Ndufs5*, *Dlat*, *Hax1*, *Mrpl15*, *Brp44l*, *4930402E16Rik/Pdpr*, and *2510003E04Rik/KBP*) and for genes involved in DNA damage response and cell-cycle checkpoints (*Chek1*, *Rad52*, *Senp8*, *BC019943/Tti2*, and *2310003H01Rik/Faap100*) (Figure 2E). To challenge the evolutionary conservation and specificity of this dual E4F1 transcriptional program, we monitored by ChIP-qPCR the presence of E4F1 at the human orthologous region of the *Dnajc19*, *Ndufs5*, *Dlat*, *Chek1*, and *Rad52* genes in human U2OS cells. This assay confirmed that E4F1 was recruited on these genes in human cells (Figure S2E).

Altogether, our data indicate that, regardless of the p53 status of the cells, E4F1 is bound to a novel regulatory motif located nearby the TSS of a limited set of genes involved in mitochondria homeostasis and stress responses, and it exerts a positive control on their expression.

all cells ($E4F1^{KO}$ and $E4F1^{WT}$) were still viable (annexin-V negative) and actively growing. The overlap defines a set of 38 genes bound and regulated by E4F1in all cell lines.

(C) MEME logo sequence analysis of the DNA fragments bound by E4F1 reveals a novel E4F1 consensus motif.

(D) ChIP-qPCR validation of E4F1 target genes in $E4F1^{WT}$ and $E4F1^{KO}$ cells 3 days after Cre-mediated inactivation of *E4F1*. Enrichments are represented as percentages of input (data are means ± SEM of four experiments).

(E) List of the 38 E4F1 direct target genes identified in (B). Table provides information about the differential expression of transcripts in E4F1$^{WT}$ and E4F1$^{KO}$ cells (ratio KO/WT), the presence of an E4F1 consensus motif near the gene promoter and its distance to TSS, and whether these genes codes for mitochondrial proteins (Mito) or for factors involved in cellular stress response and checkpoints (DDR/Chk). (See also Figure S2 and Tables S1 and S2.)

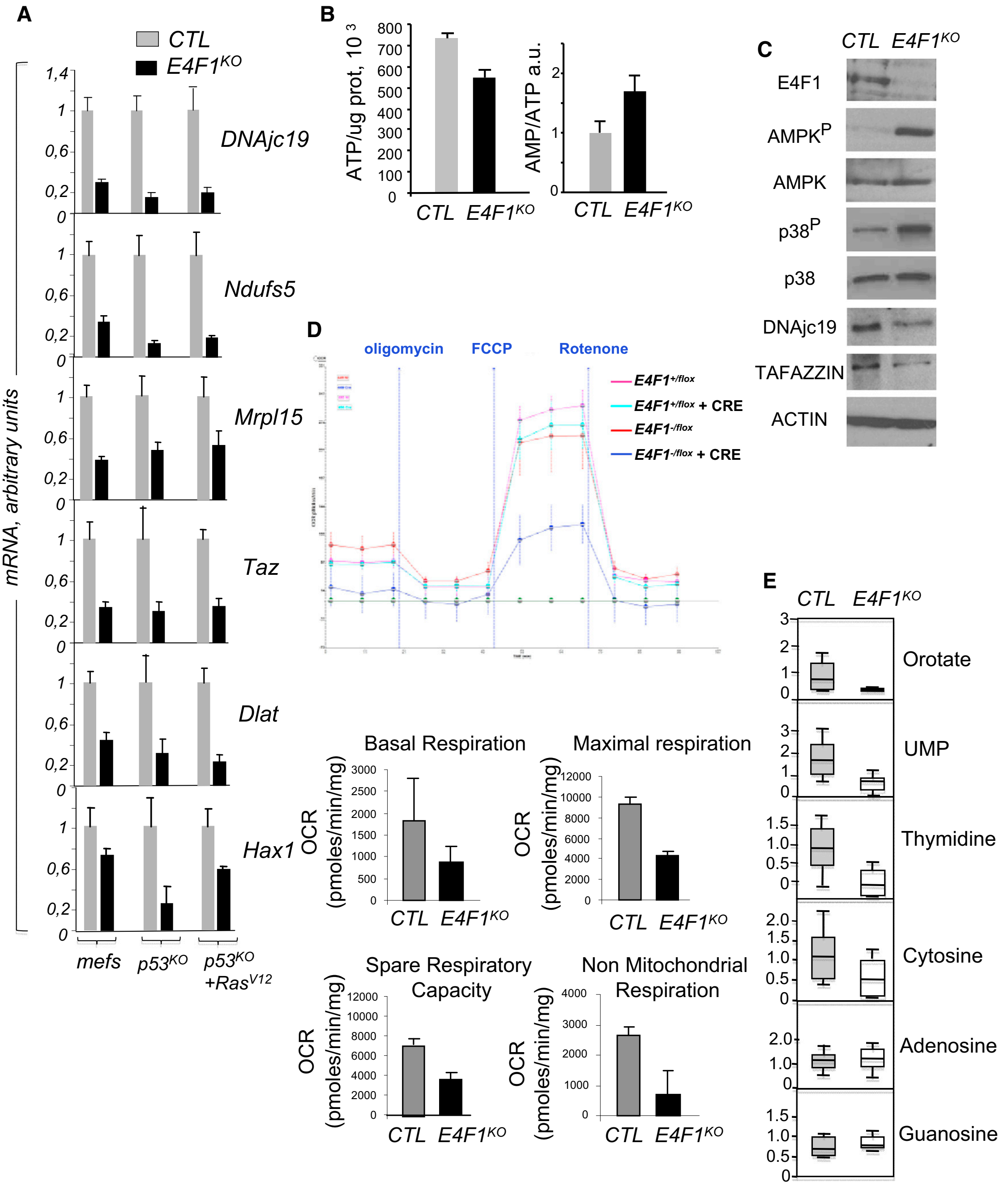

**Figure 3. E4F1 KO in Transformed MEFs Results in Mitochondria Dysfunctions that Lead to Energy Stress and Inhibition of De Novo Pyrimidine Synthesis**

(A) mRNA levels of E4F1 target genes coding for mitochondrial proteins are downregulated in E4F1$^{KO}$ cells, as measured by qRT-PCR 3 days after Cre-mediated inactivation of *E4F1* (*E4F1$^{KO}$*) in primary MEFs, *p53$^{KO}$* MEFs, or *p53$^{KO}$;HaRas$^{V12}$* transformed MEFs. In each case, corresponding Cre-treated E4F1$^{+/f}$ cells that retain an *E4F1* WT allele were used as controls (CTL). Data represent the mean ± SEM of seven experiments.

(B) ATP level (left) and AMP/ATP ratio (right) are altered in *E4F1$^{KO}$* cells, as measured 3 days after Cre-mediated inactivation of *E4F1* (*E4F1$^{KO}$*) in *p53$^{KO}$;HaRas$^{V12}$* transformed MEFs (n = 3). As in (A), corresponding Cre-treated *E4F1$^{+/f}$* transformed cells (CTL) were used as controls. Data represent the mean ± SEM of three experiments.

(C) AMPK and p38 are activated in E4F1$^{KO}$ cells. Immunoblots show the levels of AMPK, AMPK$^{P}$, p38, p38$^{P}$, DNAJC19, and TAFAZZIN on extracts from *E4F1$^{WT}$* and *E4F1$^{KO}$* transformed MEFs, treated as in (B).

(D) Representative experiment (n = 3) of an OCR performed on transformed *E4F1$^{WT}$* and *E4F1$^{KO}$* MEFs (top). Values for basal and maximal respirations, spare respiratory capacity, and nonmitochondrial respiration (evaluated upon injection of the ATP synthase inhibitor oligomycin, the uncoupling agent FCCP, and the

(legend continued on next page)

### E4F1 KO Results in Mitochondria Dysfunctions that Lead to Energy Stress and Inhibition of De Novo Pyrimidine Synthesis

The qRT-PCR validation experiments on *Dnajc19*, *Ndufs5*, *Dlat*, *Taz*, *Hax1*, and *Mrpl15* transcripts confirmed that the E4F1 mitochondrial sub-program identified by our arrays and ChIP-seq experiments was strongly downregulated in all $E4F1^{KO}$ cells (Figure 3A). These genes are involved in various biological processes and their concomitant downregulation in $E4F1^{KO}$ cells should result in mitochondrial dysfunctions, with consequences on energy production, nucleoside synthesis, or redox homeostasis, processes important for cell proliferation and survival.

We found that $E4F1^{KO}$ transformed cells showed signs of energetic stress compared to controls, as shown by the alteration in the total ATP level and the AMP/ATP ratio (Figures 3B and S3). These defects led to the induction of a p38 stress response and activation of the energy sensor AMP-activated protein kinase (AMPK) (Figure 3C). As a readout of potential mitochondrial alterations in transformed $E4F1^{KO}$ cells, we determined their oxygen consumption rate (OCR). Compared with control cells, transformed $E4F1^{KO}$ cells exhibited a strong decrease in O2 consumption (at all levels, i.e., basal respiration and maximum respiratory capacity as well as nonmitochondrial respiration, spare respiratory capacity, and ATP turnover) (Figures 3D and S3).

Mitochondrial dysfunction also has been described to entail impairment of pyrimidine de novo synthesis via a key enzyme of this pathway, the mitochondrially bound and respiratory-chain-coupled dihydroorotate dehydrogenase (DHODH). A screen for alterations in metabolites selectively affected in $E4F1^{KO}$ transformed cells revealed that *E4F1* inactivation resulted in a significantly decreased level of several intermediates in pyrimidine biosynthesis, including orotate and its downstream metabolite UMP (Figure 3E), the latter being considered as the first pyrimidine and precursor of other nucleosides. As a consequence of this downregulation in de novo UMP synthesis, the levels of the pyrimidine nucleosides, thymidine and cytosine, were also significantly decreased in $E4F1^{KO}$ cells. Unconnected to mitochondria, the neosynthesis of the purine nucleosides, guanosine and adenosine, were unaffected in these cells (Figure 3E).

### E4F1 KO Results in ROS Production, Uncompensated Oxidative Damage to Proteins, and an Accumulation of DNA Damage in Transformed MEFs

Next, we assessed whether these mitochondrial defects resulted in increased reactive oxygen species (ROS) production that can cause irreversible damage to cellular components. Using the CM-H2DCFDA probe that reacts with a broad spectrum of ROS, we confirmed that $E4F1^{KO}$ transformed cells exhibited an increase in the global level of ROS (Figure 4A). In addition, we compared the kinetics of production of superoxide by mitochondria in $E4F1^{WT}$ and $E4F1^{KO}$ cells using the mitochondria-specific and redox-sensitive dye Mitosox (Figure 4B). These analyses indicated that E4F1 depletion resulted in a significant increase in mitochondrial ROS production.

Interestingly, at least two of the E4F1 target genes, *Dnajc19* and *Taz*, have yeast orthologs that have been reported to impact on redox homeostasis (Chen et al., 2008; Mokranjac et al., 2006). *Dnajc19* codes for an inner membrane-associated J-domain protein involved in the motor machinery of the protein translocase complex that imports cytoplasmic proteins into mitochondria. *Taz* codes for the cardiolipin-modifying enzyme TAFAZZIN, which impacts on the function of multiple mitochondrial proteins requiring lipid and membrane association. Immunoblotting confirmed that DNAJC19 and TAFAZZIN protein levels were decreased in transformed $E4F1^{KO}$ MEFs (Figure 3C). Small hairpin RNA (shRNA)-mediated depletion of either DNAJC19 or TAFAZZIN in $E4F1^{WT}$ transformed cells (Figure 4D) resulted in ROS overproduction (Figure 4C) and activation of p38 and AMPK in these cells (Figure 4D). These data support the notion that the downregulation of several E4F1 targets collectively contributes to the oxidative and energetic stresses observed in $E4F1^{KO}$ cells.

We next assessed whether the burst of mitochondrial ROS detected in $E4F1^{KO}$ cells had induced detrimental oxidative damages to intracellular proteins and DNA. Protein carbonylation is an indicator of such severe oxidative damage because it is irreversible and results in proteasomal degradation. Carbonylated proteins and total proteins were labeled with hydrazide Cy5 or CyDyesTM N-hydroxysuccinimide, respectively, and then co-resolved by two-dimensional (2D) electrophoresis (Figure 4E; Baraibar et al., 2011). Abundant proteins that were differentially carbonylated between $E4F1^{KO}$ and $E4F1^{WT}$ transformed MEFs were then purified and identified by tandem mass spectrometry (MS/MS) (Figures 4F and S4A). This analysis revealed that both mitochondrial (M) and cytoplasmic (C) proteins that are known to be targeted and damaged by ROS, including VDAC2, GRP75/Mortalin, SODM, EF2, and G3P, were increasingly carbonylated upon E4F1 depletion (Figure 4F), indicative of extensive and uncompensated oxidative damages to proteins in $E4F1^{KO}$ transformed cells. High levels of ROS and decreased neo-synthesis of nucleosides both have been described as potent sources of DNA damage. Therefore, we also checked by comet assay whether $E4F1^{KO}$ cells contained DNA damage. We observed that $E4F1^{KO}$ primary MEFs exhibited increased comet assay tail moment (Figures 4G and S4B), and this defect was exacerbated in $E4F1^{KO}$ transformed MEFs. In accordance with the presence of extensive DNA damage, we detected higher levels of $\gamma$-H2AX (Figure 4H) in $E4F1^{KO}$ transformed cells.

Altogether, these data indicate that $E4F1^{KO}$ cells have accumulated uncompensated oxidative damage to proteins and extensive DNA damage.

inhibitor of the complex I rotenone) after normalization to total protein levels are shown. Vertical bars indicate the time of injection of the indicated compound. Data are represented as the mean ± SD of five wells (bottom).
(E) De novo pyrimidine synthesis is altered in $E4F1^{KO}$ cells. Metabolite concentrations were analyzed (n = 8) by LC/MS and GC/MS in $E4F1^{WT}$ (CTL) and $E4F1^{KO}$ transformed MEFs treated as in (B). (See also Figure S3.) Data represent the mean ± SEM of eight experiments.

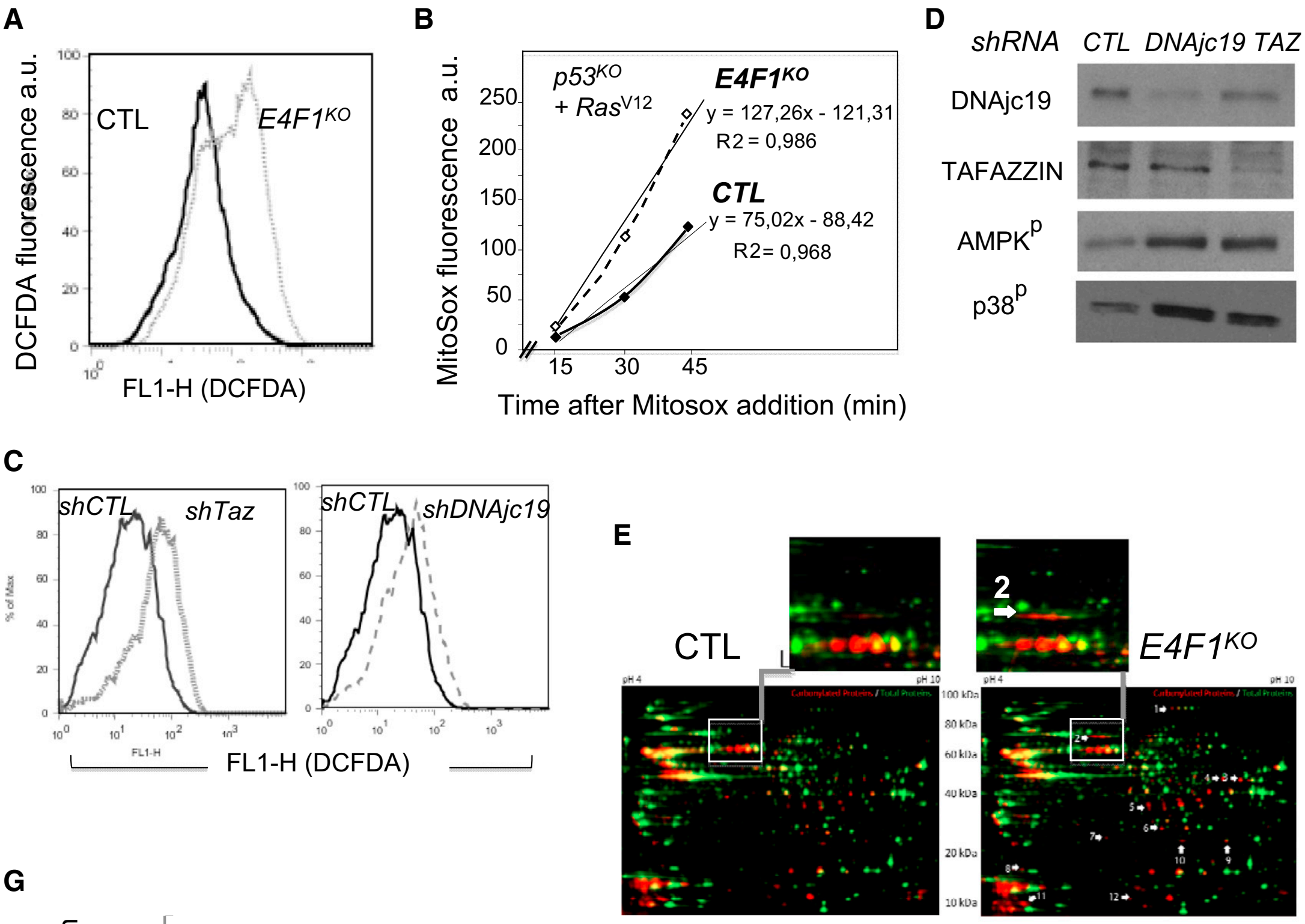

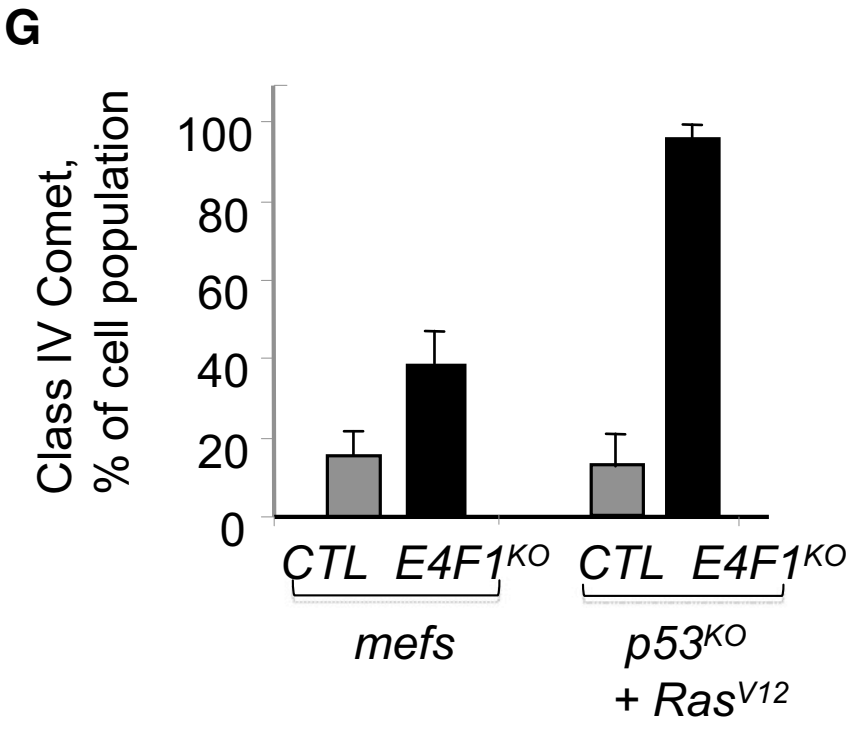

F

| Protein spot no[a] | Identified protein name | Swiss-Prot accession no[b] | RMI ratio[g] |
|---|---|---|---|
| 1 | Elongation Factor 2 | EF2_MOUSE | 4.3 |
| 2 | Stress-70 protein, mitochondrial | GRP75_MOUSE | 3.2 |
| 3 | Glyceraldehyde-3-phosphate dehydrogenase | G3P_MOUSE | 3.5 |
| 4 | Glyceraldehyde-3-phosphate dehydrogenase | G3P_MOUSE | 5.0 |
| 5 | Voltage-dependent anion-selective channel protein 2 | VDAC2_MOUSE | 5.2 |
| 6 | Glutathione S-transferase omega-1 | GSTO1_MOUSE | 2.2 |
| 7 | SAP domain-containing ribonucleoprotein | SARNP_MOUSE | 2.5 |
| 8 | Eukaryotic translation initiation factor 5A-1 | IF5A1_MOUSE | 3.7 |
| 9 | Superoxide dismutase [Mn], mitochondrial | SODM_MOUSE | 2.5 |
| 10 | Proteasome subunit beta type-2 | PSB2_MOUSE | 2.1 |
| 11 | Protein S100-A4 | S10A4_MOUSE | 2.0 |
| 12 | Cellular retinoic acid-binding protein 1 | RABP1_MOUSE | 2.1 |

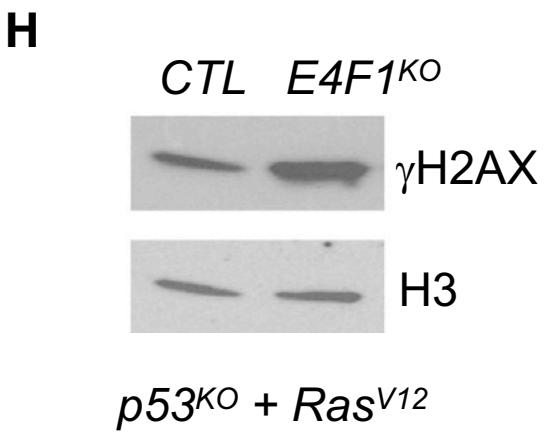

**Figure 4. Mitochondrial ROS, Uridine, and Depletion of CHK1 Contribute to the Cell Death Induced by the Inactivation of *E4F1* in p53-Deficient Transformed Cells**

(A) Cellular ROS levels are increased in *E4F1*$^{KO}$ cells. ROS levels were measured by FACScan with the fluorescent dye DCFDA in *E4F1*$^{WT}$ (CTL) and *E4F1*$^{KO}$ transformed MEFs 6 days after E4F1-mediated inactivation.

(B) Mitochondrial ROS production is increased in *E4F1*$^{KO}$ transformed cells. ROS production was measured by FACScan with the mitochondrial ROS-specific fluorescent dye Mitosox in *E4F1*$^{WT}$ (CTL) and *E4F1*$^{KO}$ transformed MEFs treated as in (A).

(C) The individual shRNA-mediated depletion of two E4F1 target genes, *Dnajc19* or *Taz*, is sufficient to generate higher levels of cellular ROS. ROS was measured with the fluorescent dye DCFDA upon transduction of *p53*$^{KO}$;*HaRas*$^{V12}$ transformed MEFs with lentiviral particles encoding shRNAs directed against DNAjc19, Taz, or an irrelevant sequence (shCTL).

(D) Immunoblotting shows the partial depletion of DNAJC19 and TAFAZZIN (Taz product) proteins and the stimulation of AMPK and p38 phosphorylation (AMPK$^{P}$ and p38$^{P}$) in cells treated by shRNAs as in (C).

(legend continued on next page)

## E4F1 Directly Controls *Chek1* Gene Expression with Impact on the CHK1-Dependent DNA Damage Response

Our list of E4F1 target genes indicated the existence of a second E4F1 transcriptional sub-program involved in DNA damage responses and cell-cycle checkpoints (Figure 2) that pointed to the *Chek1* gene. We decided to further investigate this undescribed E4F1-CHK1 pathway, considering that, together with mitochondrial dysfunctions, it could explain the extensive DNA damage and G2/M checkpoint defects observed in transformed $E4F1^{KO}$ cells (Figures 1 and 4).

Our data clearly identified E4F1 on the TSS of the *Chek1* gene in primary, $p53^{KO}$, and transformed MEFs (Figures 2D, 2E, and 5A). In contrast, E4F1 was undetectable at the *Chek2* gene that codes for the partially redundant checkpoint kinase CHK2, as illustrated in the ChIP-seq experiments (Figure 5A) and ChIP-qPCR validations (Figure 2D). Consistent with a role of E4F1 as an essential transcriptional activator of *Chek1*, expression arrays (Figure 2E) and qRT-PCR analyses (Figure 5B) showed that *Chek1* mRNA level was decreased in $E4F1^{K0}$ cells. To confirm that E4F1 was activating *Chek1* transcription, we also tested whether ectopic expression of E4F1 was able to activate a luciferase reporter construct driven by the *Chek1* promoter sequence. This reporter assay showed that full-length E4F1, but not a mutant devoided of its DNA-binding domain (ΔDBD) (Le Cam et al., 2006), stimulated this *Chek1* reporter construct in a dose-dependent manner (Figure 5C).

Importantly, CHK1 level also was decreased in $E4F1^{K0}$;$p53^{KO}$ and transformed MEFs, while *Chek2* mRNA and protein levels remained unchanged (Figures 5B and 5D). Of note, the control of E4F1 on the basal expression of *Chek1* extends beyond MEFs, since CHK1 protein downregulation also was detected by immunocytochemistry in tumor sections prepared from a previously described mouse model of histiocytic sarcomas (Hatchi et al., 2011) undergoing acute inactivation of *E4F1* in vivo (Figure 5E). In addition, we observed in human cell lines (HEL and HL60) that CHK1 level was strongly decreased upon shRNA-mediated depletion of E4F1 (Figure S5), confirming the E4F1-CHK1 connection in human cells (Figure S2E).

As previous reports showed that the CHK1-dependent checkpoint is particularly important for the survival of p53-deficient cancer cells exposed to DNA-damaging agents inducing replicative stress (McNeely et al., 2014), we next evaluated the sensitivity of p53-deficient $E4F1^{KO}$ transformed MEFs to hydroxyurea (HU). As expected, $E4F1^{WT}$ transformed control cells massively arrested in S and G2/M phases upon the addition of HU (Figure 5F). According to a major role of CHK1 in this checkpoint, inhibition of CHK1 kinase activity by the pharmacological inhibitor AZD7762 partially bypassed this cell-cycle arrest and resulted in massive cell death (Sub-G1 DNA content). Consistent with a role of E4F1 in the basal expression of CHK1, $E4F1^{KO}$ transformed cells treated with HU only exhibited a profile indicative of a compromised arrest in S and G2/M and of enhanced cell death (Sub-G1) that was similar to that of control cells treated with both HU and the CHK1 inhibitor (Figure 5F).

Altogether, these data strongly support a direct role for E4F1 in regulating CHK1 expression, with impact on the CHK1-dependent checkpoint response to DNA damage.

## The Abnormal Levels of Mitochondrial ROS, Pyrimidine Neosynthesis, and CHK1 Contribute to $E4F1^{KO}$ Cell Phenotype

Next, we addressed in rescue experiments the relative contribution of mitochondrial dysfunctions (oxidative stress and pyrimidine neosynthesis) and of CHK1 depletion in the growth arrest and cell death resulting from *E4F1* inactivation. Addition of the antioxidant ascorbate (ASC) partially bypassed the cell-cycle arrest of primary $E4F1^{KO}$ MEFs and improved, at least in the short term, their capacity to proliferate (Figures 6A and 6B). ASC also significantly decreased cell death in transformed $E4F1^{KO}$ MEFs (Figure 6C), resulting in a limited but significant restoration of proliferation (Figure 6A), observed up to 2 weeks after *E4F1* conditional inactivation (Figure S6A). Importantly, similar partial rescue of cell death was observed with mito-TEMPO, a mitochondria-targeted antioxidant (Figure 6C). Collectively, these results indicate that mitochondrial ROS is, at least in part, responsible for the proliferation defects and loss of viability of $E4F1^{KO}$ cells.

Because pyrimidine depletion due to mitochondrial defects also has been reported to induce cell death, we then tested whether the addition of uridine, a product of the pyrimidine neosynthesis pathway that bypasses the mitochondrial step of this synthesis, could also protect $E4F1^{KO}$ cells. As observed with antioxidants, uridine addition markedly decreased cell death in transformed $E4F1^{KO}$ MEFs (Figure 6D). Notably, the combination of Mito-TEMPO and uridine did not further improve the survival of these cells (data not shown).

Finally, we tested whether ectopic restoration of CHK1 protein level in $E4F1^{KO}$ transformed cells also could rescue cell death. Expression of HA-tagged CHK1 in these cells clearly enhanced their short-term survival (Figures 6E and S6B). Moreover, similar rescue was observed by transducing $E4F1^{KO}$ transformed cells with retroviruses expressing GFP-CHK1 (Figure S6C). The protective effect of CHK1 was detected up to 2 weeks after Cre-mediated inactivation of *E4F1* (Figure S6C). Noteworthy, similar CHK1 re-expression in primary MEFs did not rescue their capacity to proliferate (data not shown), likely because these

(E) Increased protein oxidation in $p53^{KO}$;$HaRas^{V12}$ transformed MEFs upon inactivation of *E4F1*. Oxiproteome (red) and total proteome (green) of $E4F1^{WT}$ (CTL) and $E4F1^{KO}$ transformed cells treated as in (A). Carbonylated proteins (red spots) and total proteins (green spots) were labeled with CyDye hydrazide and CyDye N-Hydroxysuccinimide, respectively, and then co-resolved by 2D electrophoresis. White arrows indicate protein spots increasingly oxidized (at least 2-fold, $p < 0.05$) in $E4F1^{KO}$ cells.

(F) Identification by mass spectrometry of the proteins that are increasingly carbonylated upon *E4F1* inactivation, as revealed in (A), is given.

(G) $E4F1^{KO}$ cells accumulate DNA damages. Alkaline comet assay was performed 5 days after Cre treatment of $E4F1^{-/f}$ ($E4F1^{KO}$) and $E4F1^{+/f}$ (CTL), primary or transformed MEFs ($p53^{KO}$;$HaRas^{V12}$). Histograms represent the percentage of cells with Tail moment of the fourth class (as defined in Figure S4B). (See also Figure S4.) Data represent the mean ± SEM of three experiments.

(H) γ-H2AX is activated in $E4F1^{KO}$ cells. Immunoblots show the level of γ-H2AX on extracts from $E4F1^{WT}$ and $E4F1^{KO}$ transformed MEFs.

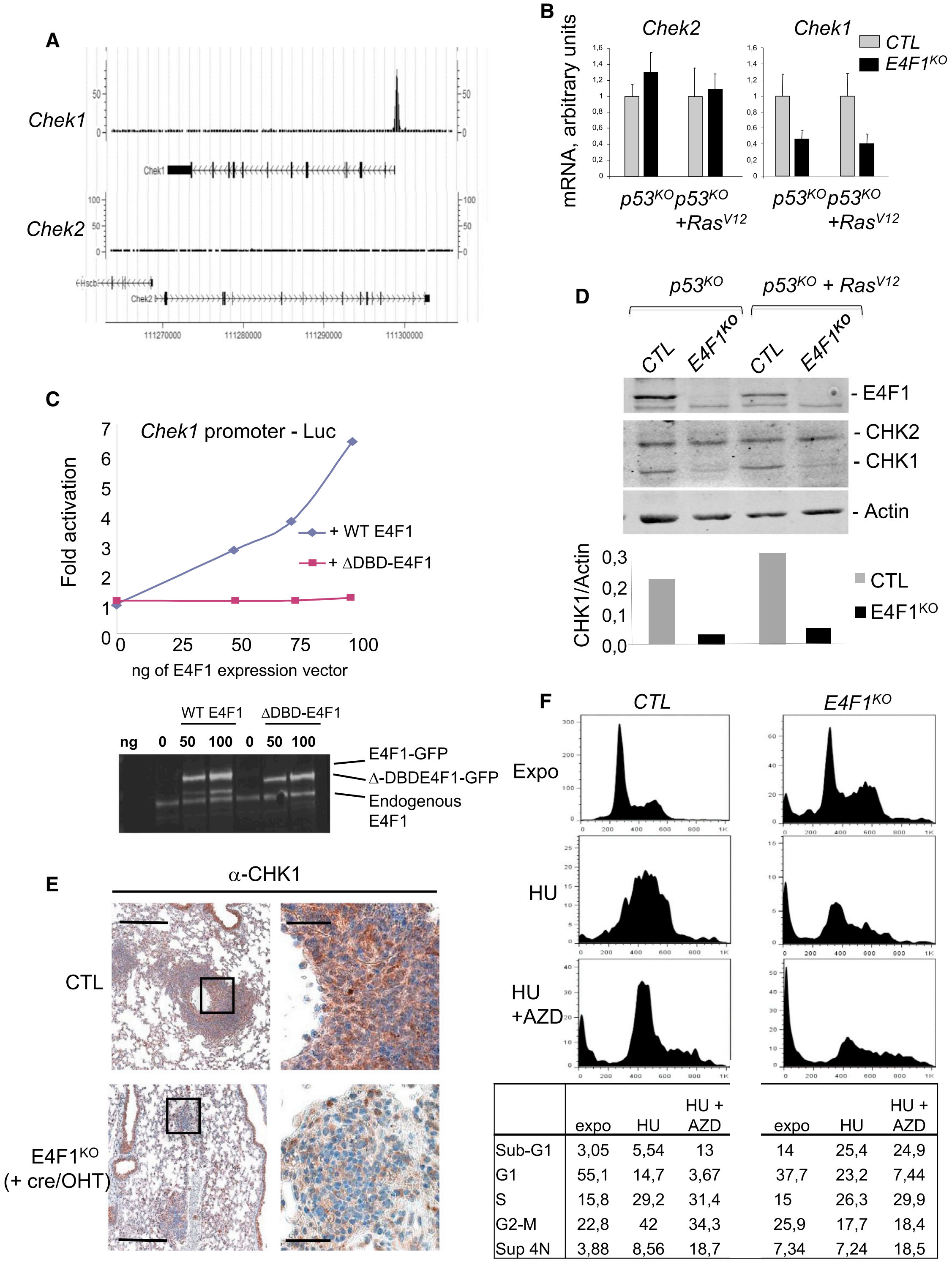

| | expo | HU | HU + AZD | expo | HU | HU + AZD |
|---|---|---|---|---|---|---|
| Sub-G1 | 3,05 | 5,54 | 13 | 14 | 25,4 | 24,9 |
| G1 | 55,1 | 14,7 | 3,67 | 37,7 | 23,2 | 7,44 |
| S | 15,8 | 29,2 | 31,4 | 15 | 26,3 | 29,9 |
| G2-M | 22,8 | 42 | 34,3 | 25,9 | 17,7 | 18,4 |
| Sup 4N | 3,88 | 8,56 | 18,7 | 7,34 | 7,24 | 18,5 |

**Figure 5. E4F1 Directly Controls *Chek1* Gene Expression with Impact on the CHK1-Dependent DNA Damage Response**

(A) E4F1 is bound to the promoter region of the *Chek1* gene, but not of the *Chek2* gene. E4F1 ChIP-seq read densities at the *Chek1* and *Chek2* genes in $p53^{KO}$;$HaRas^{V12}$ transformed MEFs are shown.

(B) Chek1 mRNA level is downregulated in $E4F1^{KO}$ cells. *Chek2* and *Chek1* transcripts were measured by qRT-PCR 3 days after Cre-mediated E4F1 inactivation ($E4F1^{KO}$) in primary $E4F1^{-/f}$;$p53^{KO}$ MEFs and in $E4F1^{-/f}$;$p53^{KO}$;$HaRas^{V12}$ transformed MEFs. Cre-treated primary and transformed $E4F1^{+/f}$ cells were used as controls (CTL). Data represent the mean ± SEM of six experiments.

*(legend continued on next page)*

cells contained p53 and were therefore still checkpoint proficient. This result is consistent with our data showing that the growth arrest of these cells mainly relies on p53 (Figure 1).

Altogether, these experiments indicate that the control of E4F1 on mitochondrial homeostasis and CHK1 expression are both required for the survival of transformed cells.

## DISCUSSION

By combining ChIP-seq and differential transcriptomic analyses in *E4F1*$^{KO}$ and *E4F1*$^{WT}$ MEFs, we show that E4F1 is bound to a limited set of genes. Characterization of E4F1-binding sites defined a new DNA motif, present nearby the TSS of these genes, that is significantly different from the one originally identified in the viral E4 promoter (Fernandes and Rooney, 1997). The decreased expression of most E4F1 target genes we identified indicate that E4F1 mainly acts as a transcriptional activator. These genes define at least two programs involved in mitochondrial homeostasis and checkpoint.

Surprisingly, one of four E4F1 target genes is coding for mitochondrial proteins involved in multiple biological processes, including components of the protein translocase (*Dnajc19*), the OXPHOS respiratory complexes (*Ndufs5*), the mitochondria transport machinery (*2510003E04Rik/Kbp1*), the pyruvate dehydrogenase (PDH) complex (*Dlat*, *4930402E16Rik*, and *Brp44l*), the cardiolipin synthetic pathway (*Taz*), an anti-apoptotic complex (*Hax-1*), and mitochondrial ribosomes (*Mrpl15*). Although E4F1 previously was described as a cell-cycle regulator, these results identify E4F1 as a novel regulator of mitochondrial homeostasis. The decreased expression of this program in *E4F1*$^{KO}$ transformed cells correlates with alterations in several biological processes that require functional mitochondria, including ATP production, pyrimidine nucleotide neosynthesis, and redox homeostasis. We believe these pleiotropic and complex mitochondrial dysfunctions result from the combined depletion of several E4F1 target genes, including Dnajc19 and Taz, the depletion of which is sufficient to recapitulate the abnormal level of ROS detected in *E4F1*$^{KO}$ cells. In addition, recent work confirms our data showing that DNAJC19 downregulation increases ROS levels (Sinha et al., 2014). Interestingly, mutations in the coding sequence of several of these E4F1 target genes (*Taz*, *DNAJC19*, *Dlat*, *Brp44l*, and *Hax1*) have been implicated in monogenic human syndromes characterized by various and complex mitochondrial dysfunctions (Schlame, 2013; Davey et al., 2006; Imbard et al., 2011; Bricker et al., 2012; Klein, 2011).

In *E4F1*$^{KO}$ cells, abnormal ROS levels and nucleotide depletion results in alterations of multiple cellular macromolecules, accumulation of uncompensated oxidative damage in cytoplasmic and mitochondrial proteins, and extensive DNA damage that eventually lead to proliferation defects or cell death. Strikingly, our data clearly indicate that E4F1 also regulates another subprogram involved in genome surveillance pathways required to cope with such damages, which includes *Chek1*, *Rad52*, *Senp8*, *BC019943/Tti2*, and *2310003H01Rik/Faap100.* How metabolism and cell-cycle control are coordinated remains poorly understood, but recent data stressed that several oncogenes or tumor suppressors impinge on both cellular processes and thereby contribute to the complex rewiring of tumor metabolism. This notion is well illustrated by the p53 pathway, where both p53 and BMI1, which were initially implicated in cell-cycle control and cell death, are now recognized as major transcriptional regulators of several metabolic pathways and mitochondria homeostasis (Liu et al., 2009; Maddocks and Vousden, 2011; Jiang et al., 2013). Similarly, E2F/RB, MYC, and OCT1 protein families recently have been shown to control genes involved in energy production, redox homeostasis, and anabolic processes (Blanchet et al., 2011; Chen and Russo, 2012; Dang, 2011). Like E4F1, they impinge on various metabolic pathways linked to mitochondrial functions (Chen and Russo, 2012). It raises important but still poorly explored questions about the overlapping, complementary functions and coordination of these metabolic and checkpoint programs.

Our rescue experiments, combined with those reported by Grote et al. in this same issue (Grote et al., 2015), indicate that the control exerted by E4F1 on CHK1 kinase occurs at multiple levels and is central for the survival functions of E4F1 in transformed MEFs and hematopoietic stem cells. Grote et al. show that E4F1 directly interacts with CHK1 protein and regulates its stability. In addition, we show that E4F1 is a potent transcriptional activator bound at the *Chek1* promoter in actively proliferating cells. Our data, together with previous reports showing the role of E2Fs (Carrassa et al., 2003; Yang et al., 2008), p53 (Gottifredi et al., 2001; Kho et al., 2004), NF-κB (Barré and Perkins, 2007), and BCL6 (Ranuncolo et al., 2008) in *Chk1* transcription, suggest that the *Chek1* promoter is an important and common hub for several oncogenic pathways. Since CHK1- and p53-dependent checkpoints are partially redundant, the upregulation of CHK1 is usually seen as the way cancer cells compensate for p53 inactivation and ensure the maintenance of a proficient checkpoint required for survival (Verlinden et al., 2007; Höglund et al., 2011). This notion is the rationale for ongoing therapeutic strategies targeting pharmacologically CHK1 kinase activity, together with genotoxic agents, in p53-deficient tumors (McNeely et al., 2014). We propose that the inactivation of

(C) E4F1 acts as a transactivator for the *Chek1* promoter. Reporter assay (top) in U2OS cells co-transfected with a luciferase reporter construct driven by *Chek1* promoter (−846 + 1,110) and increasing amounts of E4F1-GFP or of an E4F1 mutant deleted from its DNA-binding domain (ΔDBD-E4F1) is shown. Immunoblot analysis (bottom) of overexpressed E4F1-GFP proteins is shown.

(D) Immunoblotting shows that CHK1 protein level, but not CHK2 protein level, is downregulated in *E4F1*$^{KO}$ cells. Total cell extracts were prepared from cells treated as in (B).

(E) Immunocytochemistry shows CHK1 expression in *E4F1*$^{KO}$ and *E4F1*$^{WT}$ (CTL) hystiocytic sarcoma sections developed in the lung of mice transplanted with *RERT*;*E4F1*$^{+/f}$;*cdkn2a*$^{-/-}$ *or RERT*;*E4F1*$^{-/f}$;*cdkn2a*$^{-/-}$ fetal liver cells. *E4F1* was accutely inactivated in pre-formed tumors by treating mice with 4-OH tamoxifen. Bar, 1 mm (left) and 300 μm (right).

(F) *E4F1*$^{KO}$ cells have an altered response to DNA-damaging agents. DNA content analyses (PI, top) and quantifications of the Sub-G1 (dying cells), G0/G1, S, and G2/M phases of the cell cycle (bottom), in exponentially growing *E4F1*$^{KO}$ and *E4F1*$^{WT}$ (CTL) cells transformed by *p53*$^{KO}$;*HaRas*$^{V12}$ and exposed to the DNA-damaging agent HU and to CHEK1 chemical inhibitors (AZD7762), are shown. (See also Figure S5.)

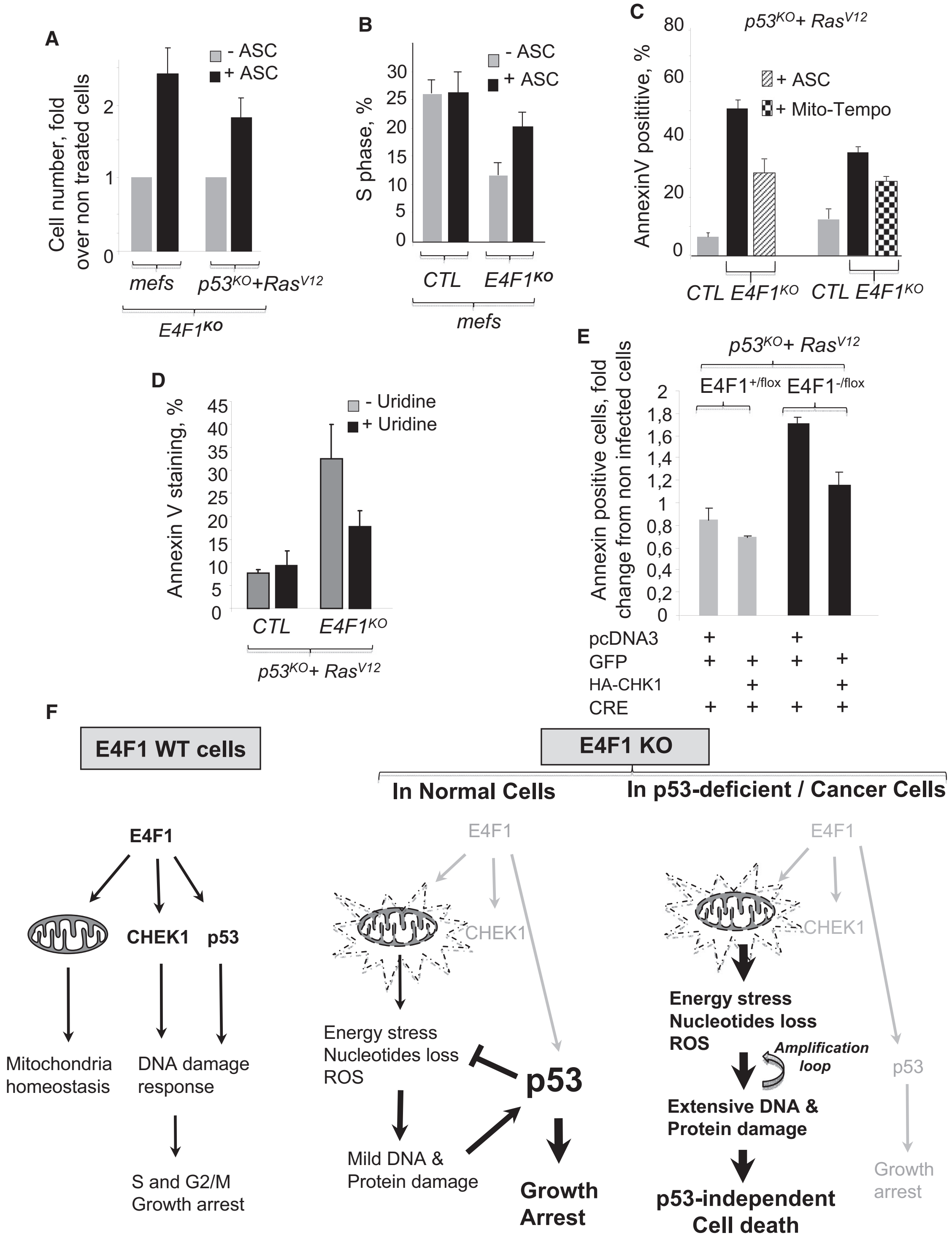

**Figure 6. Mitochondrial ROS and CHK1 Depletion Both Contribute to the Cell Death Induced by E4F1 Inactivation in Transformed Cells**

(A) Antioxidant partially restores the growth of $E4F1^{KO}$ cell populations. Proliferation curves of Cre-treated $E4F1^{-/f}$ ($E4F1^{KO}$) and $E4F1^{+/f}$ (CTL) MEFs or of $E4F1^{-/f}$;$p53^{KO}$;$HaRas^{V12}$ and $E4F1^{-/f}$;$p53^{KO}$;$HaRas^{V12}$ transformed MEFs were established as in Figure 1A, in the presence or absence of ASC (100 μM). Histograms represent the ratio of the cell numbers counted with or without ASC 10 days after Cre treatment. Data represent the mean ± SEM of three experiments.

(B) Antioxidant partially restores the capacity of $E4F1^{KO}$ primary MEFs to proliferate. BrdU incorporation was measured in Cre-treated $E4F1^{-/f}$ ($E4F1^{KO}$) and $E4F1^{+/f}$ (CTL) MEFs cultured in the presence of ASC (100 μM) 6 days after Cre-mediated inactivation of *E4F1*. Data represent the mean ± SEM of three experiments.

(legend continued on next page)

E4F1 in p53-deficient transformed cells results in multiple damages originating from mitochondria defects combined with deficiencies in both CHK1 and p53-dependent checkpoints, resulting in a deadly cocktail. Of note, the proficiency of the p53-mediated checkpoint in primary MEFs results in growth arrest and appears sufficient to limit the expansion of these damages (Figure 6F).

Our data support the notion that the E4F1-CHK1 regulatory axis is essential for the survival of p53-deficient cells, including transformed cells. This also points to a potential involvement among E4F1, CHK1, and p53, with complex feedback loops. Indeed, while we and Grote et al. (2015) reveal that E4F1 regulates positively CHK1 expression, previous reports have shown the following: (1) p53 represses *chek1* at the transcriptional level (Gottifredi et al., 2001; Kho et al., 2004); (2) CHK1 kinase regulates p53 phosphorylation (Shieh et al., 2000) and the transcriptional elongation of one of its main targets, p21 (Beckerman et al., 2009); and (3) E4F1 directly controls p53 transcriptional activities through atypical ubiquitination (Le Cam et al., 2006).

Further investigations will be required to evaluate whether the complex connections among E4F1, CHK1, and p53 are involved in embryonic and epidermal stem cells that require *E4F1* for proper homeostasis (Lacroix et al., 2010; Le Cam et al., 2004). Noteworthy, *Chek1*- and *E4F1*-deficient embryos die at a very similar developmental stage, i.e., soon after implantation (Le Cam et al., 2004; Liu et al., 2000; Takai et al., 2000).

In conclusion, our data reveal an unexpected role for E4F1 in coordinating mitochondria homeostasis and genome surveillance checkpoints. The high sensitivity of p53-deficient transformed cells to E4F1 inhibition highlights the potential interest of exploring new anti-cancer strategies targeting E4F1, or pharmacologically mimicking the deadly environment of E4F1 KO cells by combining mitochondria and checkpoint inhibitors.

## EXPERIMENTAL PROCEDURES

All experiments were approved by the University of Montpellier's Ethics Committee for Animal Welfare.

### ROS Detection, Protein Carbonylation, and 2D Map of the Oxiproteome

Total and mitochondrial ROS levels were measured by flow cytometry on live cells upon staining with CM-$H_2$DFCDA and MitoSOX probes (Invitrogen), as described previously (Hatchi et al., 2011). Kinetics of mitochondrial ROS production shown in Figure 3F are expressed as the fold increase of time-dependent changes in mean fluorescence intensity of the MitoSOX staining measured by flow cytometry (FACSCalibur, Becton Dickinson). Detection and identification of carbonylated proteins (2D Map of the oxiproteome) were performed as previously described (Baraibar et al., 2011) and as detailed in the Supplemental Experimental Procedures.

### Nucleotides Analysis and ATP/AMP Measurements

To determine the concentration of purine and pyrimidine intermediates in $p53^{KO}$;$E4F1^{WT}$;$Ha\text{-}Ras^{V12}$ and $p53^{KO}$;$E4F1^{KO}$;$Ha\text{-}Ras^{V12}$ transformed MEFs ($2 \times 10^7$ cells), soluble extracts were prepared and analyzed by gas chromatography-mass spectrometry (GC-MS) and liquid chromatography-tandem mass spectrometry (LC-MS/MS) platforms (Metabolon) on eight independent samples for each cell line. ATP and AMP concentrations (Figure 3B) were measured by high-performance liquid chromatography (HPLC) from $5 \times 10^6$ cells lysed in 65% perchloric acid. ATP levels, normalized to total protein levels, were confirmed on $2 \times 10^4$ cells using Cell Titer Glo (Promega) luminescent assay. See Supplemental Experimental Procedures for additional detail.

### ChIP-qPCR and ChIP-Seq

Detailed protocols and primer sequences used for ChIP, as well as Bioinformatic tools and parameters used to treat ChIP-seq data and annotate E4F1-bound regions, are detailed in the Supplemental Experimental Procedures.

To identify the DNA consensus sequence bound by E4F1, ChIP-seq peak DNA sequences were retrieved from bed files containing E4F1-bound region coordinates and analyzed with MEME logo suite (http://meme.nbcr.net/meme/).

### qRT-PCR and Differential Transcriptomic Analyses

To identify genes with expression levels differentially expressed between $E4F1^{WT}$ and $E4F1^{KO}$ cells, total RNA was recovered using RNeasy Mini kits (QIAGEN). Of total RNA samples, 1 μg was amplified and Cy3- and Cy5-labeled using Amino Allyl Message Amp II aRNA amplification kit (Ambion), hybridized on mouse HD12 plex Nimblegen arrays (135,000 probes/44,170 genes, platform MGX-Montpellier GenomiX, www.mgx.cnrs.fr/), and scanned on an Innoscan900 scanner (Innopsys). Raw data were normalized by Loess (Limma Package) and differentially expressed genes identified with both LIMMA and SAM packages with a false discovery rate (FDR) of 1% or 5%. Expression arrays are accessible at GEO dataset GSE57240.

For validation RT-PCR, total RNAs were isolated in TRIZOL (Invitrogen) and used to generate cDNAs by SuperScript III reverse transcription (Invitrogen) in the presence of random hexamers. Primers sequences and amplification parameters are described in the Supplemental Experimental Procedures.

## ACCESSION NUMBERS

The full series of data, including expression arrays and ChIP-seq data, reported in this paper have been deposited in NCBI's Gene Expression Omnibus and are available through accession number GEO: GSE57242 (superseries).

## SUPPLEMENTAL INFORMATION

Supplemental Information includes Supplemental Experimental Procedures, six figures, and two tables and can be found with this article online at http://dx.doi.org/10.1016/j.celrep.2015.03.024.

(C) Antioxidant partially restores viability in $E4F1^{KO}$ transformed MEFs. $E4F1^{-/f}$;$p53^{KO}$;$HaRas^{V12}$ (CTL) and $E4F1^{-/f}$;$p53^{KO}$;$HaRas^{V12}$ transformed MEFs, cultured with or without ASC or Mito-TEMPO, were probed with annexin-V-FITC 7 days after Cre-mediated inactivation of *E4F1*. Data represent the mean ± SEM of eight experiments.

(D) Uridine partially restores viability in $E4F1^{KO}$ transformed MEFs. $E4F1^{-/f}$;$p53^{KO}$;$HaRas^{V12}$ (CTL) and $E4F1^{-/f}$;$p53^{KO}$;$HaRas^{V12}$ transformed MEFs, cultured with or without uridine (50 μg/ml), were probed with annexin-V-FITC 7 days after Cre-mediated inactivation of *E4F1*. Data represent the mean ± SEM of eight experiments.

(E) Exogenous re-expression of CHK1 partially restores viability in $E4F1^{KO}$ transformed MEFs. 24 hr after Cre-treatment, $E4F1^{-/f}$;$p53^{KO}$;$HaRas^{V12}$ (CTL) or $E4F1^{-/f}$;$p53^{KO}$;$HaRas^{V12}$ transformed MEFs were co-transfected with pcDNA3-HaGFP and pcDNA3-HaChek1; 72 hr later, GFP-positive cells were probed for cell death with annexin-V-alexa568. Data represent the mean ± SEM of five experiments.

(F) Schematic representation of E4F1 functions in normal and p53-deficient or cancer cells is shown. (See also Figure S6.)

## AUTHOR CONTRIBUTIONS

G.R. and O.K. equally contributed to this work as first author. This work is a joint effort of the C.S. and L.L.C. laboratories at IRCM. G.R., L.L.C., and C.S. all equally contributed to the experiment design and writing of the paper.

## ACKNOWLEDGMENTS

This work was supported by grants from the French Ligue Nationale Contre le Cancer (LNCC, C.S. Equipe labellisée 2011), from the Agence Nationale pour la Recherche (ANR SVSE2-YinE4F1Yang2 and MetaboCycle 2 to C.S. and L.L.C.), and from the Association pour la Recherche contre le Cancer (to G.R.). Institutional support was provided by the Institut National de la Santé et de la Recherche Médicale (L.L.C.) and the Centre National de la Recherche Scientifique (C.S.), and technical support by the Montpellier Rio Imaging (MRI), Genomics (MGX) and animal facilities. T.H. and M.L. were supported by fellowships from the LNCC, and O.K. and H.D. by fellowships financed on the ANR grants JCJC-0014-01 and SVSE2-YinE4F1Yang2.

Received: May 22, 2014
Revised: December 19, 2014
Accepted: February 17, 2015
Published: April 2, 2015